\pdfoutput=1
\documentclass{JINST}
\let\ifpdf\relax

\usepackage{graphicx}
\usepackage{caption}
\usepackage{subcaption}

\title{TPB-coated Light Guides for Liquid Argon TPC Light Detection Systems}

\author{C.M. Ignarra\\
 Physics Dept., Massachusetts Institute of Technology, Cambridge, MA 02139\\
E-mail: \email{ignarra@mit.edu}}

\abstract{Light detection systems in Liquid Argon Time Projection Chambers (LArTPCs) require the detection of the 128 nm light produced during argon scintillation.  Most detectors use Tetraphenyl Butadiene (TPB) to shift the wavelength of the light into a range visible to Photomultiplier Tubes (PMTs).  These proceedings summarize characterizations of light-guides coated with a matrix of TPB in UV transmitting acrylic which are more compact than existing LArTPC light collection systems.}


\begin{document}

\section{Introduction}

LArTPCs provide high-resolution reconstruction of charged particle tracks, allowing for excellent background rejection and a precise calculation of incident neutrino energy.  This is accomplished by measuring ionization electrons produced when charged particles traverse the volume of liquid argon (LAr).  The ionization electrons are subsequently drifted by an applied electric field and measured with crossed wire planes giving reconstruction information in three dimensions, where the third dimension is given by the electron drift time.  In addition to ionization process, scintillation light is also produced as the charged particles produced in the neutrino interaction travel through the argon.  Detecting this light serves several purposes.  First, light collection provides us with timing information with few-nanosecond precision.  This is particularly important for determining the interaction time for non-beam events such a from supernova, since there is uncertainty of where the event occurred in the detector vs. when the event occurred.  Knowing the drift distance is important when calculating drift-distance-dependent effects such as charge losses and diffusion.  Light information also allows for greater background rejection, rejecting events that occur outside of the beam window but could appear to have occurred at the correct time in a different place than they actually did.  This is particularly important for surface detectors with a high flux of cosmic rays.

Future LArTPCs containing a light collection system require a slimmer profile than existing systems present in detectors such as MicroBooNE \cite{microboone} and Icarus \cite{icarus} which have 10 " PMTs behind the wire planes of the TPC.  The light in these detectors is first wavelength-shifted by a TPB coating on the PMT or on a plate directly in front of the PMT before being detected by the PMT.  This type of system and associated electronics are too bulky for future detectors, some of which propose multiple TPCs within the same volume of Liquid argon \cite{lar1,lbne}.  The system of light-guides described in Refs.  \cite{demonstration} and \cite{benchmarking} would place thin TPB coated light-guides between the anode plane assemblies, with the PMTs and electronics being located outside of the drift region.

This paper summarizes studies of light-guide materials and development of new coatings.  A detailed technical note on this work, with the full author list, is available on the arXiv \cite{benchmarking}.

\section{Components of the light-guides}

\subsection{TPB coatings} \label{sec:coatings}

TPB wavelength shifting coatings have been utilized in many neutrino and dark matter detectors aiming to detect scintillation light from liquid noble elements \cite{microboone,icarus, warp, miniclean, nEDM}.  A 1.5~$\mu$m thick evaporative coating of pure TPB has an efficiency of 1.2 $\pm$ 0.2 to absorb and re-emit 128 nm light \cite{gehman}.   This light is emitted at a peak wavelength of 425 nm, which is a good match to the peak quantum efficiency of most PMTs.  While evaporative TPB coatings have a high efficiency, they are rough and opaque where light-guides require an optically smooth surface for total internal reflection to occur.  In addition to optimizing coating efficiency, we must also optimize surface smoothness to achieve maximum attenuation length. 

The coatings discussed in this paper are as follows:

\begin{itemize}

\item Evaporative: TPB is evaporated onto a surface in a vacuum chamber. This type of coating has the highest conversion efficiency and is used in experiments such as WArP \cite{warp} and MiniCLEAN \cite{miniclean}.  The absolute efficiency of 1.5~$\mu$m thick evaporative TPB coatings has been measured as a function of wavelength in \cite{gehman}.  The evaporative samples used for the measurements in this section have a thickness of 1.87~$\mu$m.

\item PS25\% : TPB and polystyrene are dissolved in toluene in a ratio of 3:1 PS to TPB by mass and the mixture is brushed onto an acrylic light-guide where the toluene is allowed to evaporate, producing a clear film of TPB in PS.  This coating was used in Reference \cite{demonstration}.

\item UVT25\%: Similar to PS25\% except that UV transmitting acrylic is used.  UVT acrylic has a higher saturation point of TPB, enabling a ratio of 2:1 UVT acrylic to TPB by mass. This is the primary coating studied in \cite{benchmarking} and in these proceedings.

\item PS50\%: TPB and polystyrene  are dissolved in toluene in a ratio of 1:1 by mass.  This amount of TPB over-saturates the PS and crystalizes out of solution onto the surface of the PS, leading to a higher efficiency than PS25\% or UVT33\%.  This type of coating is more robust and less expensive than evaporative coatings and is used in the MicroBooNE experiment \cite{microboone}, but like the evaporative coatings, it is too rough for light-guiding.

\end{itemize}

\begin{figure}
\centering
\includegraphics[width=0.5\textwidth]{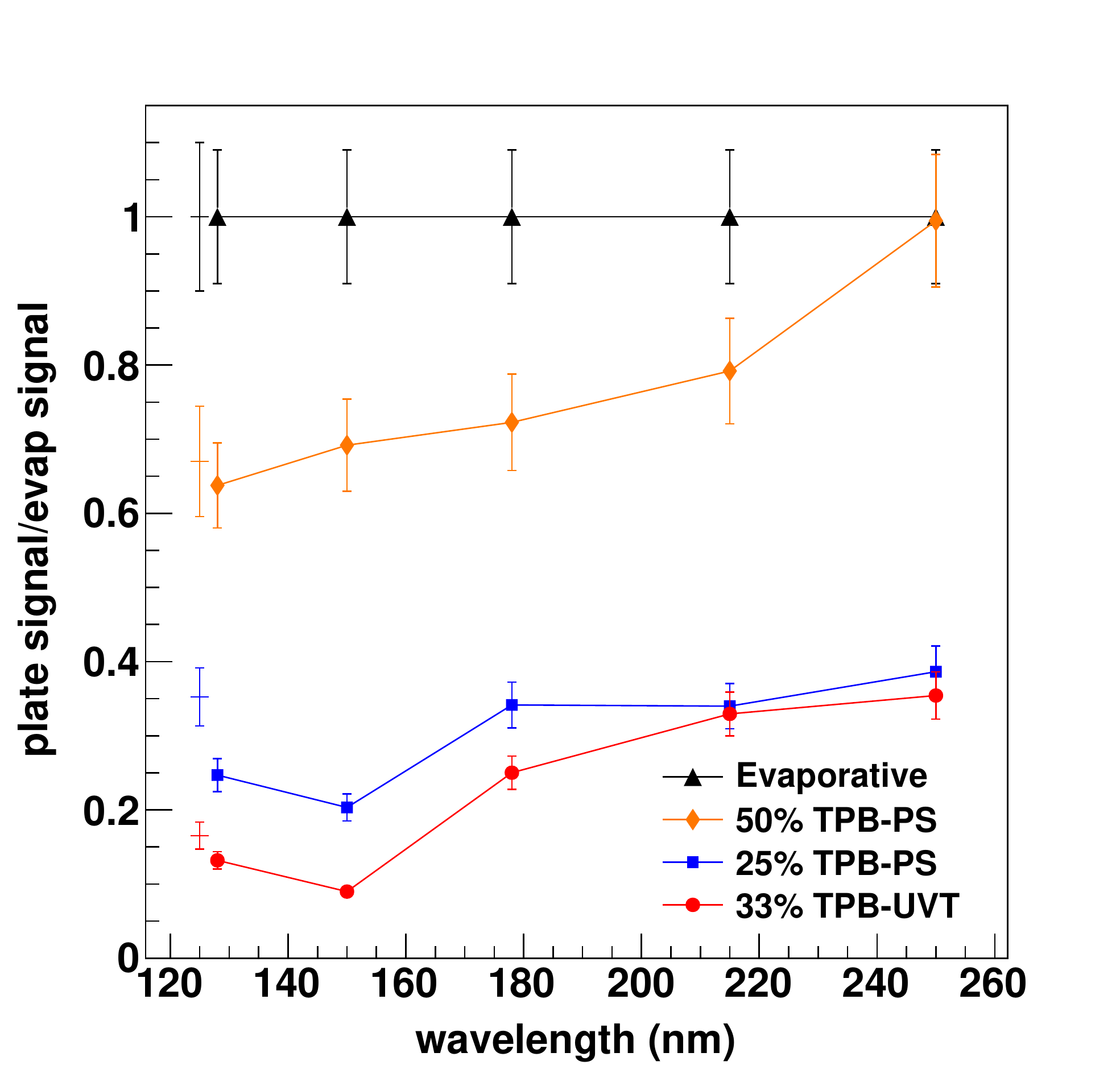}
\caption{{\bf Connected points:} Coating efficiency measured in a vacuum monochrometer 
as a function of wavelength normalized to measurements of evaporative samples.
 {\bf Left-most points:} Tests in LAr using an alpha source which are at 128nm and over-plotted onto the vacuum measurements for comparison to the 128 nm vacuum point.  Errors on the evaporative samples must  also be taken into account when comparing LAr with vacuum for a particular coating, but not when comparing one type of coating with another in the same medium.
\label{fig:coatings}}
\end{figure}

The coating efficiencies were determined by applying the coating to a plate and measuring the light yield transmitted through the plate.  They were tested in a McPherson 234 vacuum monochrometer at wavelengths between 128 nm and 250 nm and also in LAr at 128 nm.  The results are shown in Figure \ref{fig:coatings}.  The vacuum setup consists of light from a deuterium source incident on a grating which isolates a previously selected wavelength to send to the sample chamber.   A PMT is located outside of the vacuum region on the opposite side of the sample in order to detect the light after it is shifted and has passed through the coating.  The results are normalized to the evaporative coating measurement in order to remove the spectrum of the deuterium lamp.  Error bars on bottom three lines only take into account 
the errors associated with those measurements, 
allowing for comparison amongst the samples,
while error bars on the evaporative sample illustrate errors on the evaporative measurement which also affect the ratio values of the other points.

The liquid argon measurements are shown in the same plot to the left of the 128 nm vacuum point.  They were made using a $^{210}Po$ alpha source which produces 128 nm light via argon scintillation and is located directly in front of the sample with a Hamamatsu R7725mod 2-inch PMT directly behind the sample.  These measurements are also normalized to an evaporative coated sample in order to compare to previous results, though deviations within errors of either evaporative measurements would cause the other samples to shift as well which must be taking into account when comparing the results. Taking this into account, one can see that the datasets in vacuum and argon are in fairly good agreement.  We have previously seen visible mechanical degradation of evaporative coatings due to the violent boiling of the argon around the sample during submersion, so this could account for the the sample-to-evaporative ratios in LAr being high relative to the vacuum measurements.  The vacuum measurements were taken using evaporative samples that had never been exposed to LAr.

All samples from the coatings measurements described in this section and the light-guide measurements in the following sections were carefully stored in a dark environment proceeding (and in-between) measurements to mitigate the damaging effects of light \cite{degradation, benzophenone}.

\subsection{Base acrylic \label{sec:airattenuation}}

\begin{figure}
\centering
\includegraphics[width=0.6\textwidth]{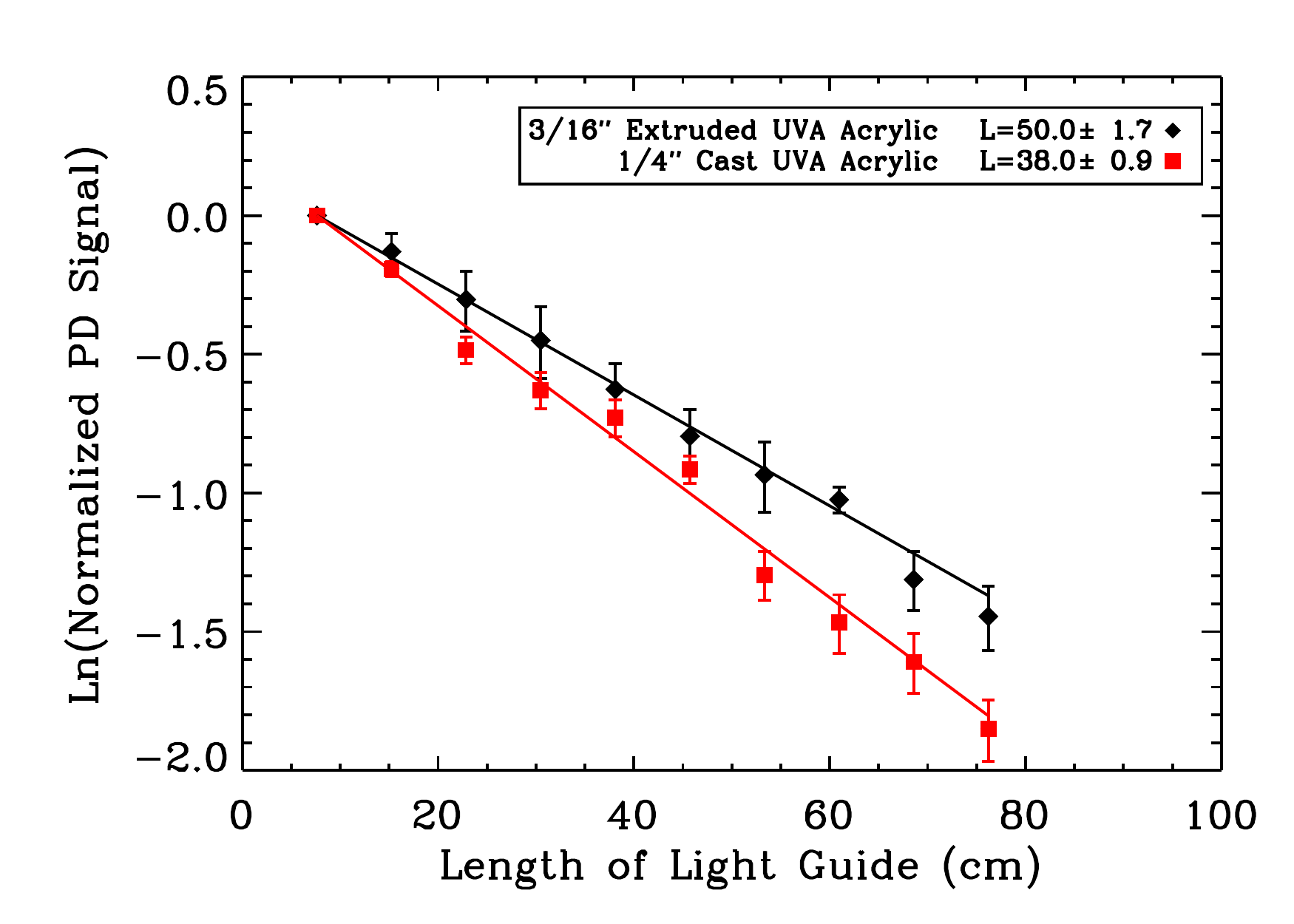}
\caption{Measurements of the attenuation length in air of extruded and cast acrylic bars performed at 420 nm result in an attenuation length of 50$\pm$2~cm for the extruded acrylic and 38$\pm$1~cm for cast UVA acrylic.}
\label{fig:airattenuation}
\end{figure}

For these tests, we used cast acrylic from McMaster Carr since it has a longer bulk attenuation than the extruded acrylic of reference \cite{demonstration}. However, measurements done by S. Mufson and B. Baptista at Indiana University of the uncoated attenuation length in air (Figure \ref{fig:airattenuation}) yield significantly shorter attenuation lengths than the bulk attenuation lengths expected from most commercial acrylics.  These are generally the order of several meters.  This indicates that surface effects dominate the attenuation losses, as is described in further detail in Ref \cite{bensimulation}.  Further information on this attenuation length measurement can be found in Ref. \cite{benchmarking}. We will attempt to improve attenuation lengths by further polishing of our base acrylic to minimize surface effects in a future study.

\section{Measurements of light-guides in LAr}

\subsection{Setup}

The test stand uses a cryogenic Hamamatsu R7725mod 2-inch PMT which is situated such that it will be in contact with the light-guide under study once submerged in LAr.  A $^{210}Po$ alpha source is placed at a distance of 5 mm from the light-guide and emits 5.3 MeV $\alpha$ particles which produce 128 nm argon scintillation light.  This short path length is chosen to as to minimize the effects of photon absorption due to dissolved impurities \cite{n2}. The apparatus is submerged in liquid argon in a glass dewar and data is acquired using an Alazar Tech ATS9870 digitizer.  More information on the test setup can be found in references \cite{demonstration} and \cite{benchmarking}.

\subsection{Analysis}

\begin{figure}
\centering
\includegraphics[width=0.7\textwidth]{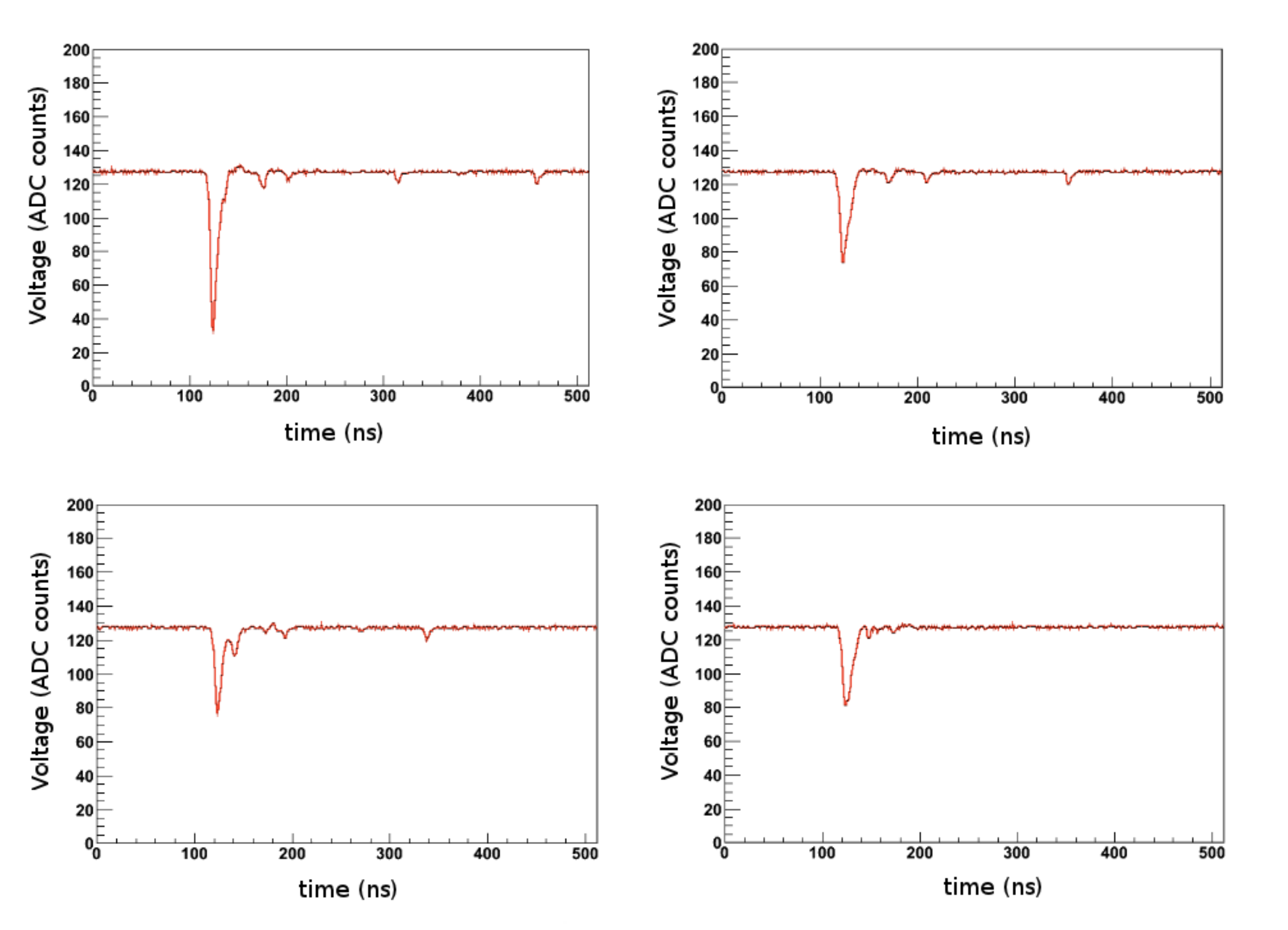}
\caption{Examples of events read out with a waveform digitizer on a scale of $\pm$200~mV/256~ADC counts. 
The trigger pulse consisting mainly of early light and single-p.e. pulses due to the late light are clearly visible in each event. 
\label{fig:pulses}}
\end{figure}

Scintillation light in Liquid Argon has two components: a fast component ($\tau_{1/2} =  6~ns$) and a slow component ($\tau_{1/2} =  1.6~ \mu s$) \cite{icarusscintillation}, corresponding to light emission from an initial singlet or triplet state.  Both processes produce light at 128 nm.  Due to quenching of the late light component by impurities in the argon \cite{WARPN,WARPO}, the light in the trigger pulse primarily consists of light from the fast scintillation path.   Despite the quenching, some late light is still seen and single photons from this process can be seen following the trigger pulse.  Example waveforms consisting both of the trigger pulse and trailing late-light pulses are shown in Figure \ref{fig:pulses}.  Because most of the late light pulses consist of one photoelectron, we can use a distribution of late-light pulses in order to calibrate the gain of the PMT. 

Measurements are made by integrating over the pulse from 30 ns before the trigger to 120 ns after the trigger, where the trigger is set at a pulse height of 17 ADC counts.

\begin{figure}
\centering
\includegraphics[trim=0cm 2cm 0cm 2cm, clip=true, width=0.6\textwidth]{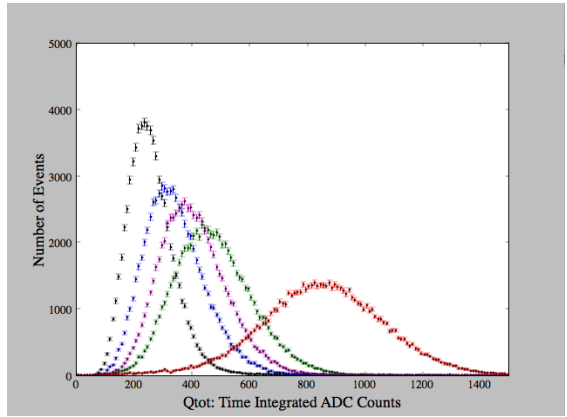}
\caption{ 
A variety of example integrated charge distributions at various distances along a cast acrylic bar with a UVT33\% coating. In order from left to right, these distances are: Black: 50 cm,  blue: 40 cm, magenta: 30 cm, green: 20 cm, red: 10 cm. 
\label{fig:qtotexamples}}
\end{figure}

 Examples of measurements of these integrated charge distributions at various distances along a cast acrylic bar with a UVT33\% coating can be seen in Figure \ref{fig:qtotexamples}.  

\subsection{Results}

\begin{figure*}
\centering
\mbox{\includegraphics[width=2.4in]{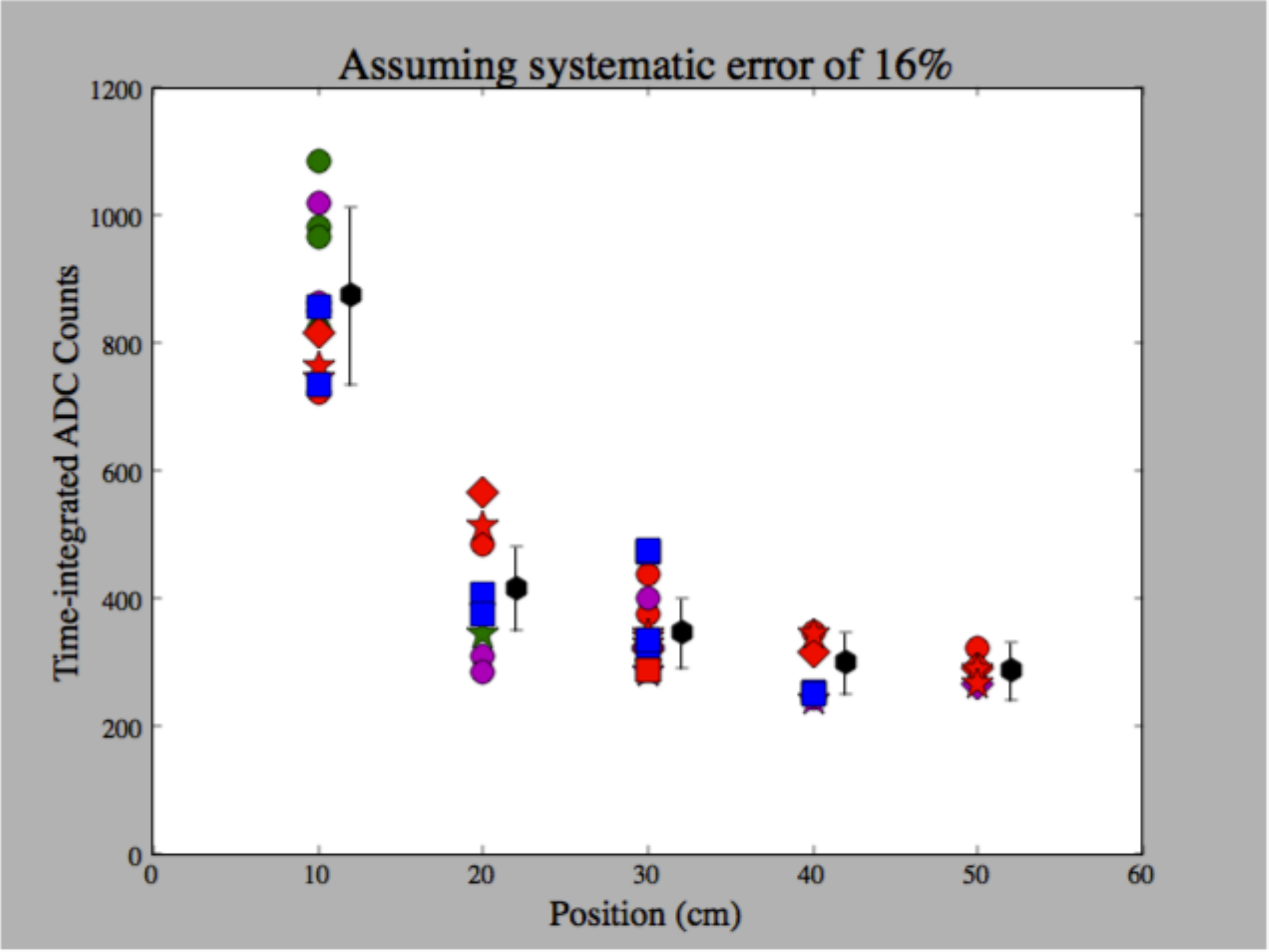}}
\qquad
\mbox{\includegraphics[width=2.4in]{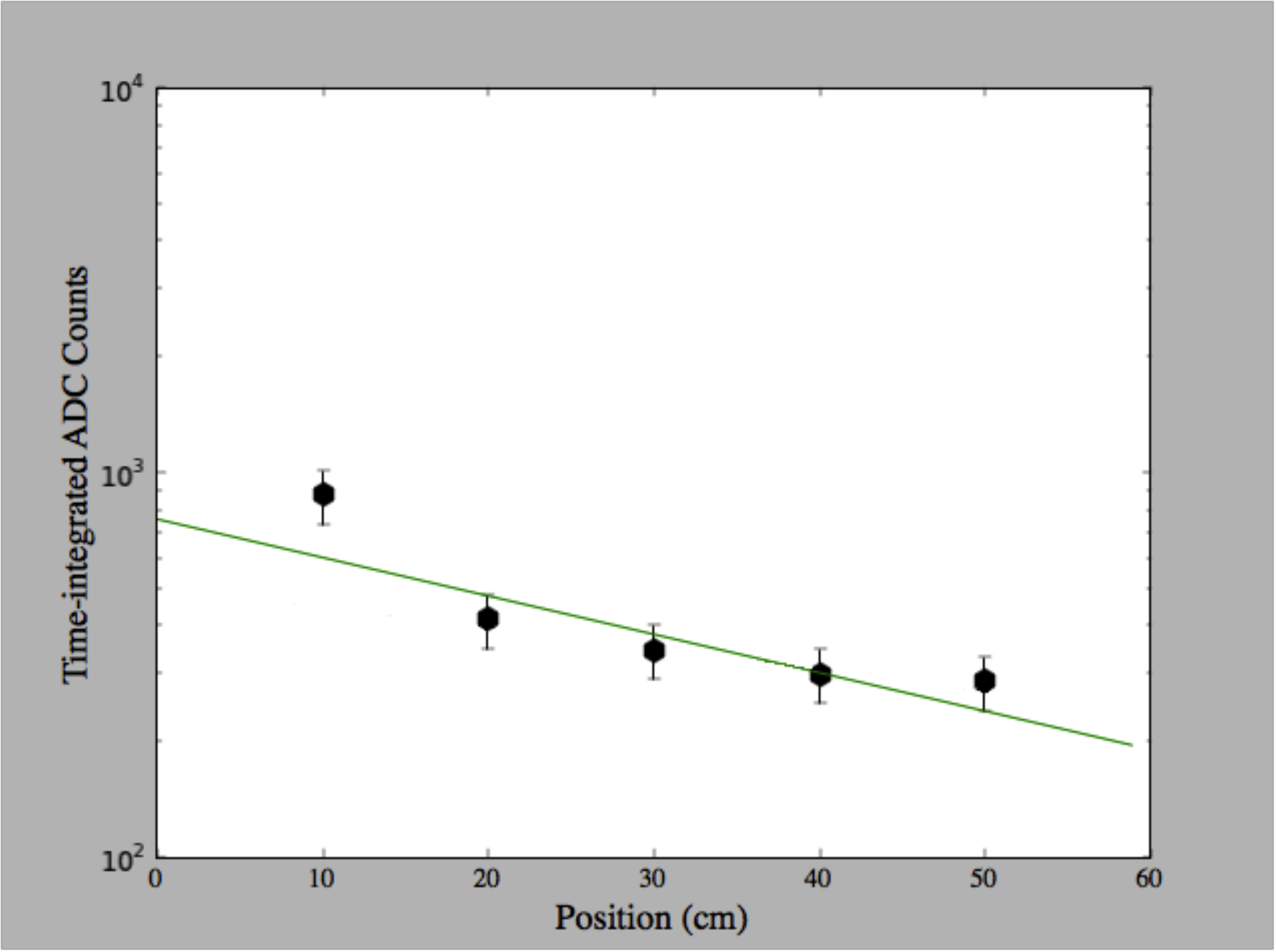}}

\caption{Attenuation length measurement of cast acrylic bars with UVT25\% coating.  Left: colors and symbols represent four batches of LAr and and four bars respectively, showing consistency when changing these variables.  Right:  Exponential fits to the mean measurement at each position yields an attenuation length of 44cm. \label{fig:larattenuation}}

\end{figure*}

Figure \ref{fig:larattenuation} shows the results of averaging measurements such as those in  Figure \ref{fig:qtotexamples} over four bars and four batches of LAr taken during a period of 60~days.    Measurements at each position were averaged for a measurement of the attenuation length.   

The attenuation length of cast acrylic bars with a UVT 25\% coating was measured to be 44 cm, (compared with 38 cm in air) though the result does not look particularly exponential, which has lead to the simulation studies described in Ref \cite{bensimulation}.  Ref  \cite{bensimulation} uses the measured attenuation in air in order to calculate a surface absorption coefficient to predict attenuation behavior in argon, which results in a better match to our data than the exponential fit shown.

Similar measurements of light-guides with a PS25\% coating show a reduced light yield with a shorter attenuation length \cite{demonstration, benchmarking}.  Further investigation is necessary to understand why the total efficiency of UVT25\% coatings on light-guides exceed that of PS25\% coatings, despite the higher conversion efficiency of the PS25\% coating when the coating was isolated in Section \ref{sec:coatings}.  We hypothesize that the better index of refraction match and/or better coefficient of expansion match of the UVT acrylic to the base acrylic may increase the amount of light captured by the light-guide.

\section{Conclusion \label{sec:conclusion}}

We have characterized a design for a thin-profile light-guide system for light collection in LArTPCs and provided a comparison between materials and coatings for this system.     We have improved our light yield by a factor of 3 (as measured at 10 cm from the source) from our previous studies of Ref \cite{demonstration} and are continuing work to further improve our coating efficiency.  Significant improvements in attenuation length of the base acrylic are expected to be possible by attempting to eliminate surface losses such as through polishing or by moving to the acrylic suggested by Ref. \cite{stuart} which has recently been shown to have a higher attenuation length for this application.

\section*{Acknowledgments}

The author thanks the National Science Foundation (PHY-1205175) for their financial support

\bibliographystyle{unsrt}

\bibliography{ignarra_lidine_proceedings}

\end{document}